\documentclass[%
 aip,
 apl,%
 amsmath,amssymb,
reprint,%
]{revtex4-1}
\usepackage[T1]{fontenc}
\usepackage[utf8]{inputenc}
\usepackage{amsmath,amssymb}
\usepackage{color}
\usepackage{graphicx}
\usepackage{amssymb}

\newcommand{\beqa}{\begin{eqnarray}}
\newcommand{\eeqa}{\end{eqnarray}}
\newcommand{\beq}{\begin{equation}}
\newcommand{\eeq}{\end{equation}}

\newcommand{\cH}{\mathcal{H}}

\begin{document}
\title{Trigonal distortion of topologically confined channels in bilayer Graphene.}
\author{A. S. N\'u\~{n}ez}
\email{alnunez@dfi.uchile.cl}
\affiliation{Departamento de F\'{i}sica, Facultad de Ciencias F\'{\i}sicas y Matem\'aticas, Universidad de Chile, Blanco 
Encalada 2008, Santiago, Chile}
\author{E. Su\'{a}rez Morell}
\author{P. Vargas}
\affiliation{Departamento de F\'{i}sica, Universidad T\'{e}cnica
Federico Santa Mar\'{i}a, Casilla 110-V, Valpara\'{i}so Chile}

\begin{abstract}
In this work we show that the trigonal warping of the electronic bands in bilayer graphene 
dramatically modifies the behavior of the topologically confined one-dimensional modes  
due to an inhomogeneous bias that changes sign across a channel.
Up to four zero modes are present, depending on the orientation of the channel, these zero modes are related with a fractionalization of the topological charge due to the trigonal warping.
\end{abstract}

\maketitle

Intervalley transport and coherence in graphene-based systems is an appealing subject, due to both the intriguing physical phenomena that it entails and the great potential that it displays from an applied perspective. Among others, valley-dependent resistance effects and valley-filter devices,  have been proposed in several contexts\cite{Xiao2007,Schomerus2010}. Importantly, basic theoretical estimates prove that such devices are well within current or soon to come technological capabilities.

Bilayer graphene (BLG), two coupled graphene layers, has an electronic structure remarkably different from graphene. 
Unlike single layer graphene (SLG) the electronic dispersion near the K points is parabolic and a gap can be induced by applying an external electric 
field\cite{key-7} perpendicular to the layers. 

The coherence between carriers on each of the layers of BLG opens up an additional effect. It has been shown \cite{key-Topo1} that biased BLG is characterized by non-trivial topological structure associated with the behavior of the pseudo-spin degree of freedom.  This topological property leads to several fascinating predictions.\cite{key-Topo2,key-Topo3,PhysRevLett.104.216406}  In particular it can be shown that  confined gapless bands can be induced in BLG  by  applying an inhomogeneous bias that changes sign across a channel. In this way a one dimensional channel populated with chiral zero modes is created with potential applications. It can be done for instance in a graphene p-n junction\cite{Williams03082007}, provided that the transition is sharp enough.
The existence of such states in the regions of space that are in the midst of a change of topological charges has been extensively studied and form part of the physics of many systems. 

At the heart of the matter lies the fact that, under the effect of the transverse electric field,  the pseudo-spinor eigenfunction warps  the Brillouin zone in a non-trivial way enclosing it an integer number of times. This number is a topological charge (the so-called Chern number) that cannot be untied in a continous fashion. Correspondingly changes in this topological charge are bound to be discontinuous and zero energy states are located nearby such discontinuities.The topological protection of the states together with their different effects from  aforementioned effect has been suggested as a plausible valley-filter device. 

In the present work we include in the behavior of the chiral bands the effects of trigonal warping. We show that such distortion of the electronic bands in bilayer graphene dramatically modifies the behavior of the one-dimensional confined modes due to an inhomogeneous bias that changes sign across a channel. We have found up to four zero modes per valley per spin depending on the angle and the width of the domain wall. The effect is enhanced in wide domain walls however in such a case more branches can be found inside the gap. These other branches are localized within the domain wall and are of the same nature that the ones obtained in modelling a contact of two materials with inverted bands\cite{Inverted-Pankratov}.They are present regardless of the trigonal warping. 

A low energy reduced (dimensionless) Hamiltonian for the stacked AB bilayer, including the interaction responsible for the trigonal warping and an external field can be written as\cite{key-7,key-Topo5}:
\beq
\cH= \varepsilon_{0} \left( \begin{array}{cc}   
V & k_+^{2}+ s k_- \\
k_-^{2}+ s k_+ & -V \\
\end{array} \right)
\label{eq:1}
\eeq
where $\varepsilon_{0}=(\gamma_{3}/\gamma_{0})^{2}\gamma_{1}\approx 4 $ meV,  the units of $k$ are $k_{0}=2 \gamma_{3} \gamma_{1}/(\sqrt{3}a \gamma_{0}^{2})$, where we have used $\gamma_{0}=3.16$ eV, $\gamma_{1}=0.39$ eV,
and $\gamma_{3}=0.315$ eV, and $a=0.246$ nm is the lattice constant. Additionally we employed the artificial parameter $s$  to include continuously the effect of trigonal warping from $s=0$, where it is strictly absent, to $s=1$ where it acquires its full value. The symbol $k_\pm$ stands for $\xi k_x\pm i k_y$, where  $\xi=1$ for $\cH$ around $K'$ and $\xi=-1$ for $K$.
We can also write the Hamiltonian in the following manner ready to assess topological invariants\cite{key-Topo4}
 \beqa
\cH &=& \varepsilon_0 \; \textbf{g}(k_{x},k_{y}) \cdot \sigma, 
\eeqa
where $\textbf{g}(k_{x},k_{y})=\textbf{g}_{0}(k_{x},k_{y})+s \textbf{g}_{w}(k_{x},k_{y})$. In this equation 
$\textbf{g}_{0}(k_{x},k_{y})=(k_{x}^{2}-k_{y}^{2},2\xi k_{x}k_{y},V)$
 and $\textbf{g}_{w}(k_{x},k_{y})=(-k_{x},k_{y},0)$
where $\sigma$ are the Pauli matrices. $\textbf{g}_{w}$ is the part responsible for the trigonal warping.\\
The topological invariant $N_{3}$ which describes K points is the winding number of the mapping of the sphere $\sigma_{2}$ 
around the K points to the 2-sphere of the unit vector $\hat{\textbf{g}}=\textbf{g}/\vert\textbf{g}\vert$ is\cite{key-Topo4}:

\begin{equation}
N_{3}=\frac{1}{4\pi} \int d^{2}k \: \hat{\textbf{g}} \cdot \left[   \frac{\partial\hat{\textbf{g}}}{\partial{k_x}} \times \frac{\partial\hat{\textbf{g}}}{ \partial{k_y}}\right] 
\label{eq:2}
\end{equation}
The topological charge $N_{3}$ can be found with no difficulties from the equations given above. It turns out that $N_3=\xi\; {\rm sign}(V)$, independently of $s$. Despite this invariance the effects of trigonal warping are far from trivial. The trigonal warping splits the Fermi point, $K$ into four pockets: three Dirac points away from the $K$ point with charge $+\xi$ and the $K$ 
point with charge $-\xi$.\cite{key-7,key-Topo6}\\
A reversal of the bias voltage over a region of space is therefore associated with a change of $|\delta N_3|=2$
for each valley and spin. This  gives the minimum number of zero-energy left movers minus right movers\cite{key-Topo4}, there should be then at least two branches per valley per spin that cross the Fermi-level.

\begin{figure}[htbp] 
   \centering
   \includegraphics[scale=0.9]{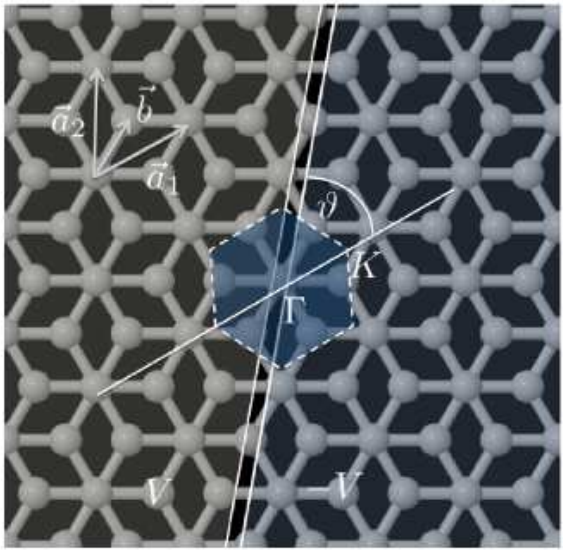} 
   \caption{A transverse bias voltage with a domain wall that makes an angle $\vartheta$ with the $\Gamma-K$ axis. Propagative states are localized around the domain wall. In the figure the circular spots represent carbon atoms arranged in a lattice with trigonal symmetry as in BLG. The hexagon represents the first Brillouin zone of BLG with the $\Gamma$ point centered.  The domain wall, between transverse potential $V$ and $-V$ has its geometry referred to the $\Gamma-K$ axis, additionally, its width is denoted as $\ell$. }
   \label{fig:example}
\end{figure}

We next proceed to model the electronic states confined by a bias-domain wall oriented along an axis $\hat{n}$ that makes an angle $\vartheta$ from the $\Gamma-K$ axis.
For that purpose we consider a potential of the form $\varepsilon_0 V(x)=\varepsilon_0V_0 \tanh(x_\perp/\ell)$ where $x_\perp$ is the coordinate perpendicular to the $\hat{n}$ axis,$\ell$ is given in $a$ (lattice constant) units, $V_0$ is an adimensional measure of the strength of the potential and
$\varepsilon_0$ is defined below Eq. (\ref{eq:1}).
The domain wall therefore connects two regions of space with different topological characters. The change in the Chern number as one crosses the wall is 2. According to the index theorem, this  gives the number of zero-energy left movers minus right movers\cite{key-Topo4},  there are at least two branches per valley per spin that crosses the Fermi-level.
In order to study the detailed structure of those zero energy bands, we will in the following diagonalize the Hamiltonian. The diagonalization of the Hamiltonian  can be accomplished by noting that the momentum along the axis defined by the domain wall, $\hat{n}$,  is a good quantum number and states can be labeled by it, $p_y$. We then rotate the axis by an angle $\vartheta$ such that the $y$-axis lies along the domain wall. The Schr\"odinger equation can be transformed into:
\begin{equation}
\left(i \partial_x +\mathcal{A}\right)^2\Psi +\mathcal{B}\Psi=\left(\varepsilon-V(x)\sigma_z\right)\sigma_x\Psi\label{eq: Hamiltonian 1D}
\end{equation}
where $\mathcal{A}=i \xi (p_y\sigma_z+1/2\; s {\rm e}^{i3\xi\vartheta \sigma_z})$ and $\mathcal{B}=1/4\; s^2 {\rm e}^{i6\xi\vartheta \sigma_z}$. 
The special case of $s=0$ (neglecting the effects of trigonal warping) can be solved analytically in terms of hypergeometric functions.\cite{key-Topo3} However, the inclusion of trigonal warping makes it more convenient to pursue a numerical approach.  

\begin{figure}[htbp] 
   \centering
   \includegraphics[angle=270]{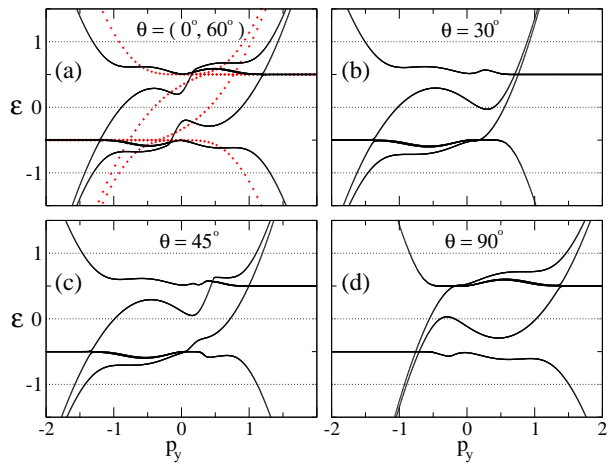}  
   \caption{One dimensional channels band structure for different angles of the domain wall. Trigonal distortion is seen to induce several effects on the band structure. All the plots  are referred to the $K^\prime$ valley.( $V_0=0.5$ and $l=1$). In $(a)$ the bands for $s=0$ are included. Energies are given in $\varepsilon_{0}$  units.}
   \label{fig:1Dbands}
\end{figure}

\begin{figure}[htbp] 
   \centering
   \includegraphics[scale=1,angle=270]{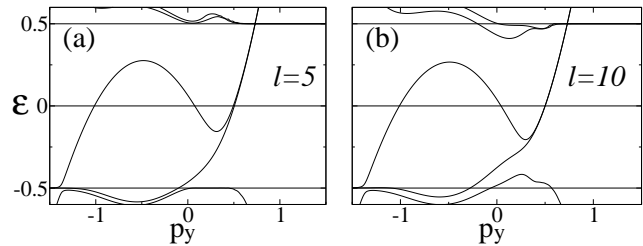}  
   \caption{Band structures for different widths of the domain wall and for an angle of $30^{\rm o}$. The four zero modes are evident and the other branches inside the gap in $(b)$. Energies and $\ell$ are given in  $\varepsilon_{0}$ and $a$ units respectively}
   \label{fig:zeromodes}
\end{figure}

The resulting eigenproblem can be solved independently for each value of $p_y$. For a given value of $p_y$ we discretize the axes across the wall and perform a numerical diagonalization of the resulting Hamiltonian.
The results are listed in  Fig. (\ref{fig:1Dbands}) for several values of the angle $\vartheta$ ($V_0=0.5$). It is evident that despite the severe deformation suffered by the bands, there are still two bands that cross the Fermi level in each case. The velocity is heavily distorted and the trigonal distortion induces local energy extrema that are not present in the isotropic case. The location and depth of those valleys formed by trigonal distortion depend strongly on the angle that the wall makes with the lattice. \\
We found that for an angle of $30^{\rm o}$ ($90^{\rm o}$ for the other branch) the distortion is maximum, and is more evident from Fig. (\ref{fig:zeromodes}) that one of the confined bands crosses three times the zero energy. There is an angle ($\sim 23^{\rm o}(37^{\rm o})$) where two crossings also can be found making a total of three zero modes. The effect is enhanced by increasing the domain wall width.
This interesting regularity can be found from the picture of fractionalization of Chern numbers.  The integrand in eq. \ref{eq:2} (Berry Curvature) depends on the applied bias, the result is always the same but four well defined peaks can be found in the limit of small bias, one central dip with a contribution of $-1/2$ and three others with a $+1/2$ value each, farther from the center and narrower the lower the bias. The total value of the integral is still +1 and  on the other side of the wall, the values are opposite. \\

The index theorem sets an inferior bound to the number of zero modes in ($N_{3}^{a}-N_{3}^{b}$). If we sum the sign of the slopes at the crossings we obtain the number predicted by the theorem.
Rotation of the domain wall breaks the symmetry between the two regions in such a way that the effective Chern numbers ($N_{3}^{a,b}$) take different values on both sides of the wall. For instance $(3\times(+1/2)-1/2)$ - $(3\times(-1/2)+1/2)$ gives two, but the number three can be obtained with $(3\times(+1/2)+1/2-1/2)-(3\times(-1/2))$ or shifting the $\pm1/2$ to the other side, and the four zero modes can be formed with $(3\times(+1/2)+1/2) - (3\times(-1/2)-1/2)$. Such reorganization of the topological charge on each side of the wall lead us to the proper number of zero modes. It is not possible to do a combination to obtain more than four zero modes with those fractions. These half integer contribution has been explored theoretically recently in several models\cite{half1,half2}.\\
There are three contributions to the appearance of these zero modes, at low bias, important to split the topological charge, second an angle of the domain wall around $30^{\rm o} (90^{\rm o})$, breaking the symmetry, and a wide domain wall with a wide wave function and in turn a k vector more localized, the latter needed to take account of the slight splitting of the K point. \\ 
 Another important effect that can be appreciated is the lack of electron-hole symmetry in the system except in the case $\vartheta=0^{\rm o}$ ($\sim 60^{\rm o}$.) This effect is expected since the Hamiltonian in Eq.(\ref{eq: Hamiltonian 1D}) fails to commute with the charge conjugation operator.

\begin{figure}[htbp] 
   \centering
   \includegraphics[scale=0.65] {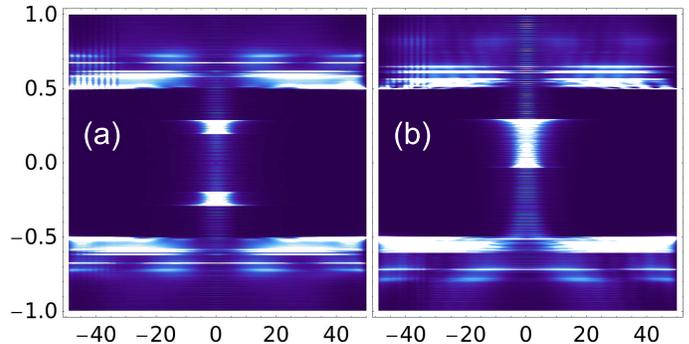}  
   \caption{Local density of states, $A(E,x)$, depicted as with a color map (arbitrary units) as function of \textbf{$\varepsilon$} (vertical axis in $\varepsilon_{0}$ units) and $x$ (horizontal axis in a units)($V_0=0.5$ and $l=1$). In (a) $\vartheta=0^{\rm o}$ and (b) $\vartheta=30^{\rm o}$  . It is evident from these plots that the propagating states that cross the Fermi energy are located within the domain wall.}
   \label{fig:LDOS}
\end{figure}

The propagating modes need to lie in the channel defined by the domain wall, otherwise their energy will be gapped.
In Fig. \ref{fig:LDOS} it is clear that the singularities associated with the distortion of the confined channels lie within the gap and they are well localized inside the domain wall, singularities are different for different angles as expected from the band structure, we expect an increase in the conductivity with respect to the isotropic case\cite{Barbier13122010} in the ballistic regime, due to the appearance of the extra zero mode.\\

We have studied the distortion that trigonal warping exerts over the topologically confined modes in transverse bias-domain walls in BLG. We have shown that, despite their topological robustness, the band structure of those modes are heavily altered by the anisotropy induced by trigonal warping; depending on the angle and on the width of the domain wall up to four zero modes can be found, they can be explained in terms of a fractionalization and reorganization of the topological charge.

{\em Acknowledgments}
ESM thanks L. Brey and L. Chico for helpful discussion.
ASN work was partially funded by Proyecto Fondecyt 11070271 Proyecto Basal FB0807-CEDENNA and NUCLEO MILENIO NºP10-061-F. ESM acknowledges UTFSM for the internal grant PIIC-DGIP and Proyecto Basal FB0807-CEDENNA. PV acknowledges support from FONDECYT grant 1100508 and CEDENNA, Chile.




\end{document}